\documentclass[aps,prl,superscriptaddress,twocolumn,amsmath,longbibliography]{revtex4-2}

\usepackage{graphicx, color}
\usepackage{xcolor}
\usepackage{amsfonts,amssymb,amsmath}
\usepackage{siunitx}
\usepackage{bm}
\usepackage{bbold}
\usepackage{braket}
\usepackage{mathtools}
\usepackage{dsfont}
\usepackage{lineno}
\usepackage{bigints}
\usepackage[normalem]{ulem}
\usepackage{enumitem}
\usepackage{amsfonts}

\begin{document}

\title{Parallel-Field Hall effect in ZrTe$_5$}

\author{Yongjian Wang}
\affiliation{Physics Institute II, University of Cologne, Z\"ulpicher Str. 77, 50937 K\"oln, Germany}

\author{Thomas B\"omerich}
\affiliation{Institute for Theoretical Physics, University of Cologne, Z\"ulpicher Str. 77, 50937 K\"oln, Germany}

\author{A. A. Taskin}
\affiliation{Physics Institute II, University of Cologne, Z\"ulpicher Str. 77, 50937 K\"oln, Germany}

\author{Achim Rosch}
\affiliation{Institute for Theoretical Physics, University of Cologne, Z\"ulpicher Str. 77, 50937 K\"oln, Germany}

\author{Yoichi Ando}
\email[]{ando@ph2.uni-koeln.de}
\affiliation{Physics Institute II, University of Cologne, Z\"ulpicher Str. 77, 50937 K\"oln, Germany}

\begin{abstract}
Parallel-field Hall effect is the appearance of a Hall voltage $V_{\rm H}$ that is transverse to the current $I$ when the magnetic field $B$ is applied parallel to $I$ (i.e. $B \parallel I \perp V_{\rm H}$). Such an effect is symmetry forbidden in most cases and hence is very unusual. Interestingly, the existence of a finite parallel-field Hall effect was reported for the layered topological semimetal ZrTe$_5$ and was proposed to be due to Berry curvature. However, it is forbidden for the known symmetry of ZrTe$_5$ and the possible existence of a misaligned out-of-plane magnetic field was not completely ruled out. Here, we elucidate the existence of the parallel-field Hall effect in ZrTe$_5$ with careful magnetic-field alignment. We interpret this result to originate from symmetry breaking and quantitatively explain the observed parallel-field Hall signal by considering a tilting of the Fermi surface allowed by broken symmetry. 

\end{abstract}
\maketitle

Recently, the Hall effect is attracting renewed attention \cite{Nagaosa2010, Armitage2018, Ando2013, Xiao2010, Liang2018, Mutch2021, Ma2019, Du2021, Zhou2022, Ge2020, Manna2018, Gao2021}, because it often reflects a topological nature of the energy bands. A nonzero integrated Berry curvature of the bands gives rise to the intrinsic anomalous Hall effect (also called topological Hall effect) that is observed not only in the ordinal Hall configuration (i.e. the current $I$, the magnetic field $B$, and the Hall voltage $V_{\rm H}$ are all orthogonal to each other) but also in the in-plane configuration (i.e. $V_{\rm H}$ is measured in the plane spanned by $I$ and $B$). 
Because the Lorenz force is always perpendicular to the magnetic field, it has been argued that the observed signal in the in-plane configuration is likely of unconventional origin such as the Berry curvature. Here we call it unconventional Hall effect (UHE).

It is important to make a distinction between the conventional planar Hall effect (PHE) and the in-plane UHE. The PHE is not really a Hall effect but is due to a resistivity anisotropy, which gives rise to a finite $V_{\rm H}$ when $I$ is in-between the anisotropy axes \cite{Taskin2017, Burkov2017, Nandy2017}. It is easy to tell between the two, because the PHE is symmetric with $B$, while UHE is antisymmetric with $B$. Hence, the $V_{\rm H}$ signal which is antisymmetric with $B$ and is observed in the in-plane configuration is unusual and it is a useful probe of the symmetry and topology of the system. 
We note that the antisymmetric in-plane UHE was historically called ``longitudinal Hall effect'' \cite{Grabner1960} and it was explained as a higher-order effect \cite{Grabner1960, Baranskii1966} or a two-band effect \cite{Bauhofer1985} in conventional semiconductors.
In this context, an extreme case is the UHE when $I$ and $B$ are parallel to each other, i.e. $I \parallel B \perp V_{\rm H}$. In this configuration, a finite UHE is forbidden in almost any symmetry classes \cite{Wang2024} and even the PHE should disappear \cite{Taskin2017, Burkov2017, Nandy2017}. When a finite signal is observed in such a configuration, it is very unusual and suggests a significant symmetry breaking. Although the $I \parallel B$ configuration is often called longitudinal, we call such a Hall effect ``parallel-field Hall effect'' (PFHE) here to avoid possible confusion with the historical longitudinal Hall effect \cite{Grabner1960}. 

The semimetal ZrTe$_5$ is one of the first materials where the antisymmetric UHE was found for in-plane magnetic fields, along with a pronounced anomalous Hall effect for out-of-plane magnetic fields \cite{Liang2018}. The transport properties of ZrTe$_5$ is dominated by the three-dimensional (3D) Dirac cone located at the $\Gamma$ point, which endows a topological character to this material \cite{Weng2014, RYChen2015, Li2016, Liu2016, YZhang2017, HWang2018, Liang2018, Li2018, Xu2018, Tang2019, Zhang2019, Sun2020, Fu2020, Galeski2021, Wang2022, Wang2023, Wang2024}. In ZrTe$_5$, the antisymmetric Hall signal is observed for almost any direction of $B$ when $I$ is along the $a$-axis and $V_{\rm H}$ is measured in the cleavage ($ac$) plane, but it is the strongest for the in-plane configuration (i.e. $B$ is in the $ac$ plane) \cite{Liang2018}. 
Since such a behavior can be naturally understood if the Berry curvature vector is pointing to the $b$-axis, it was proposed that the observed UHE is due to a finite integrated Berry curvature coming from a splitting of the 3D Dirac cone into Weyl cones \cite{Liang2018}. Later, it was theoretically shown that Weyl cones are not necessary for the in-plane UHE but a finite anomalous orbital polarizability is needed instead \cite{Wang2024}. A finite anomalous orbital polarizability requires a lower crystal symmetry than $Cmcm$ that is generally assumed for ZrTe$_5$ \cite{Weng2014, RYChen2015}, but the recent observation of a magnetochiral anisotropy (MCA) indicates that inversion symmetry must be broken in ZrTe$_5$ \cite{Wang2022} and the correct symmetry is most likely $Cm$ or lower, which allows for the in-plane UHE \cite{Wang2024}.

It is worth noting that even with the relatively low $Cm$ symmetry, the PFHE is symmetry forbidden for $B \parallel a$-axis \cite{Wang2024}. 
This can be easily understood as follows: Due to a $bc$ mirror plane included in $Cm$, a $bc$ mirror transformation maps $I$ to $-I$ for a current along the $a$-axis without affecting a parallel magnetic field (because $B$ is a pseudovector) and the voltage in $c$ direction. Thus, a Hall signal that is odd in $I$ cannot occur.
Nevertheless, a recent paper reported a nonzero PFHE for $B \parallel a$-axis in ZrTe$_5$ and interpreted it to be due to a pair of tilted Weyl cones \cite{Ge2020}. 
This paper \cite{Ge2020} argues that the relevant tilting can be obtained from a model respecting the $Cm$ symmetry. This seems, however, not possible based on our
symmetry analysis. On the experimental side, in Ref. \cite{Ge2020} the in-plane Hall effect was measured by rotating the magnetic field in the $ac$ plane, for which it is very difficult to guarantee that the magnetic field was always exactly in-plane. Even a small misalignment of the magnetic field out of the $ac$ plane produces a sizeable $V_{\rm H}$ in ZrTe$_5$, because the dependence of the anomalous Hall signal on the out-of-plane magnetic field $B_{\perp}$ is very unusual in ZrTe$_5$, taking a maximum at small $B_{\perp}$ and showing an abrupt sign change across $B_{\perp}$ = 0 \cite{Mutch2021}. Therefore, a very careful alignment of the magnetic field to guarantee $B_{\perp}$ = 0 would be necessary for unambiguously elucidating a nonzero PFHE in ZrTe$_5$. 

In this work, we performed such a careful measurement of the in-plane Hall effect in ZrTe$_5$ and found that the PFHE is indeed nonzero, although the effect is smaller than that reported in Ref. \cite{Ge2020} (see supplement \cite{SM} for actual comparison). This is surprising, because such a signal is forbidden in ZrTe$_5$ if it respects the $Cm$ symmetry. Our result thus points to a further symmetry breaking in ZrTe$_5$ at low temperature. To understand the observed antisymmetric PFHE, we present a model which considers a slight rotation of the Fermi surface with respect to the crystalline lattice, which could be caused by an additional interaction in the absence of symmetry. This model can quantitatively explain the result, supporting that the antisymmetric PFHE in ZrTe$_5$ is a result of low symmetry. Our analysis further demonstrates that antisymmetric UHE can be observed without the contribution of the Berry curvature when the symmetry is sufficiently low, which is a useful insight for future studies of topological materials.
    
\begin{figure}[t]
	\centering
	\includegraphics[width=8.6cm]{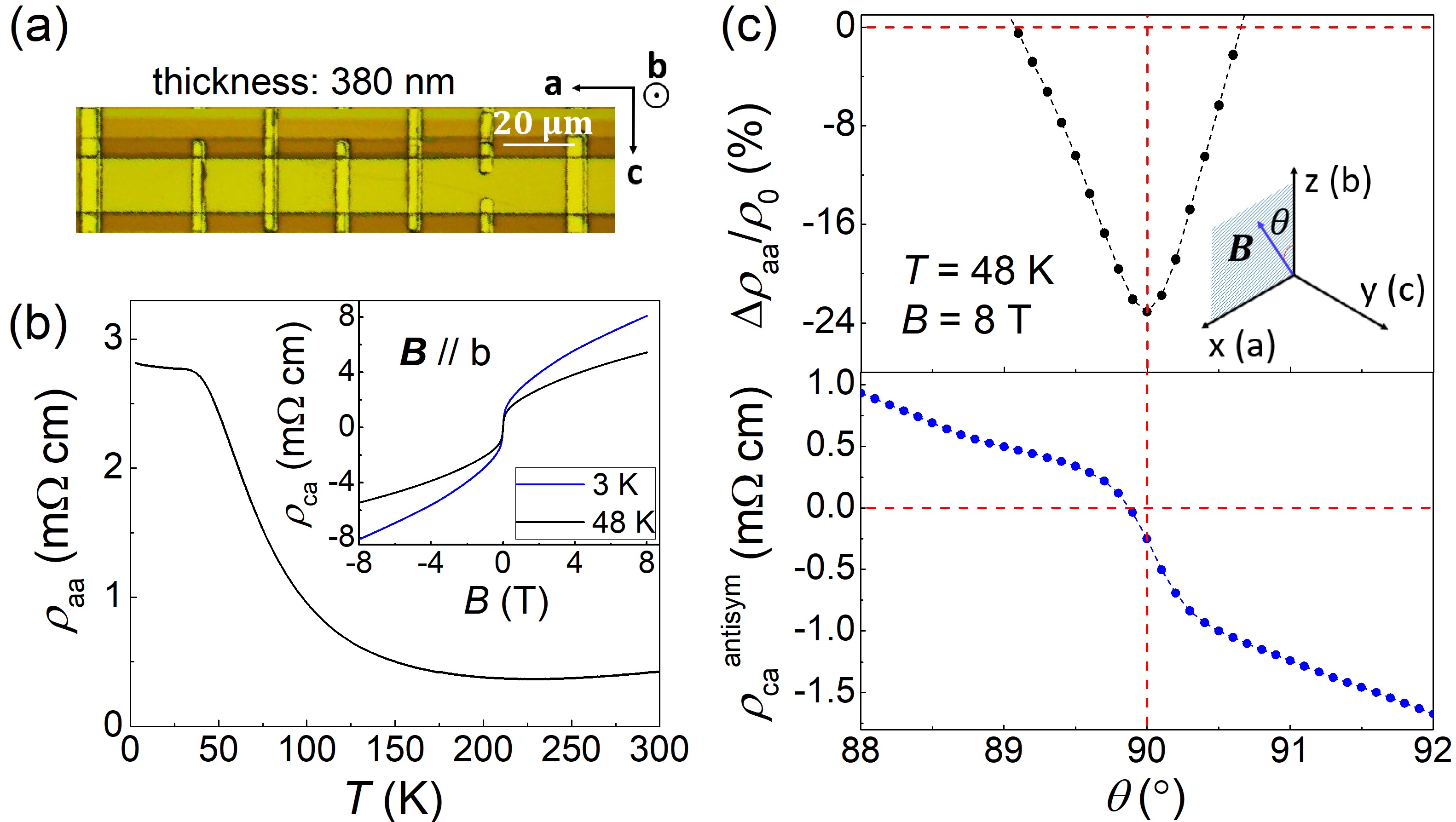}
	\caption{(a) Photograph of the micro-device of ZrTe$_5$ used in this study. (b) Temperature dependence of $\rho_{aa}$. Inset: $\rho_{ca}$ vs $B$ for $B \parallel b$ at 3 K and 48 K showing the anomalous Hall effect. (c) Dependences of the magnetoresistance and the antisymmetric component of $\rho_{ca}$ at 8 T on the magnetic-field angle $\theta$ in the $ab$ plane measured at 48 K. Inset shows the definition of $\theta$. }
	\label{fig:1}
\end{figure}

Bulk single crystals of ZrTe$_5$ are soft and easily bent. Hence, when a bulk crystal is used for the magnetotransport measurements, there is always a possibility that the crystals is slightly bent and the magnetic field is locally misaligned even when the global alignment is optimized. We therefore fabricated a micro-flake-based device as shown in Fig. 1(a) using a lithography technique \cite{SM}. The resistivity-peak temperature $T_p$ of the device was 0~K as shown in Fig. 1(b). The single crystals were grown in the same way as that described in Ref. \cite{Wang2022} to obtain such $T_p$ = 0~K samples. The long axis, along which the current flows, is $a$ and the flake plane is $ac$. Note that it is the convention of ZrTe$_5$ to take the out-of-plane direction as $b$-axis, such that the $x-y-z$ coordinate corresponds to $a-c-b$. The anomalous Hall effect \cite{Liang2018} was reproduced as shown in the inset of Fig. 1(b). 

To realize the configuration $I \parallel B \parallel a$-axis with zero $B_{\perp}$ as best as possible, we measured the longitudinal resistivity $\rho_{aa}$ and the Hall resistivity $\rho_{ca}$ simultaneously upon changing the magnetic-field angle $\theta$ in the $ab$ plane to cross the $a$-axis, and identified the sharp minimum in $\rho_{aa}$ to correspond to $B_{\perp}$ = 0 (i.e. $\theta$ = 90$^{\circ}$). As one can see in the upper panel of Fig. 1(c) and in Ref. \cite{SM}, the accuracy of the alignment with this method is better than 0.1$^{\circ}$. The lower panel of Fig. 1(c) shows that the antisymmetrized $\rho_{ca}$ is clearly nonzero at $\theta$ = 90$^{\circ}$, which indicates that the antisymmetric PFHE is unambiguously presented in ZrTe$_5$. 
Note that this measurement was performed at 48 K, which was chosen to avoid possible complications coming from the nonlinear transport \cite{Wang2023} and MCA effect \cite{Wang2022} that have been observed at lower temperature. 

\begin{figure}[t]
	\centering
	\includegraphics[width=8.6cm]{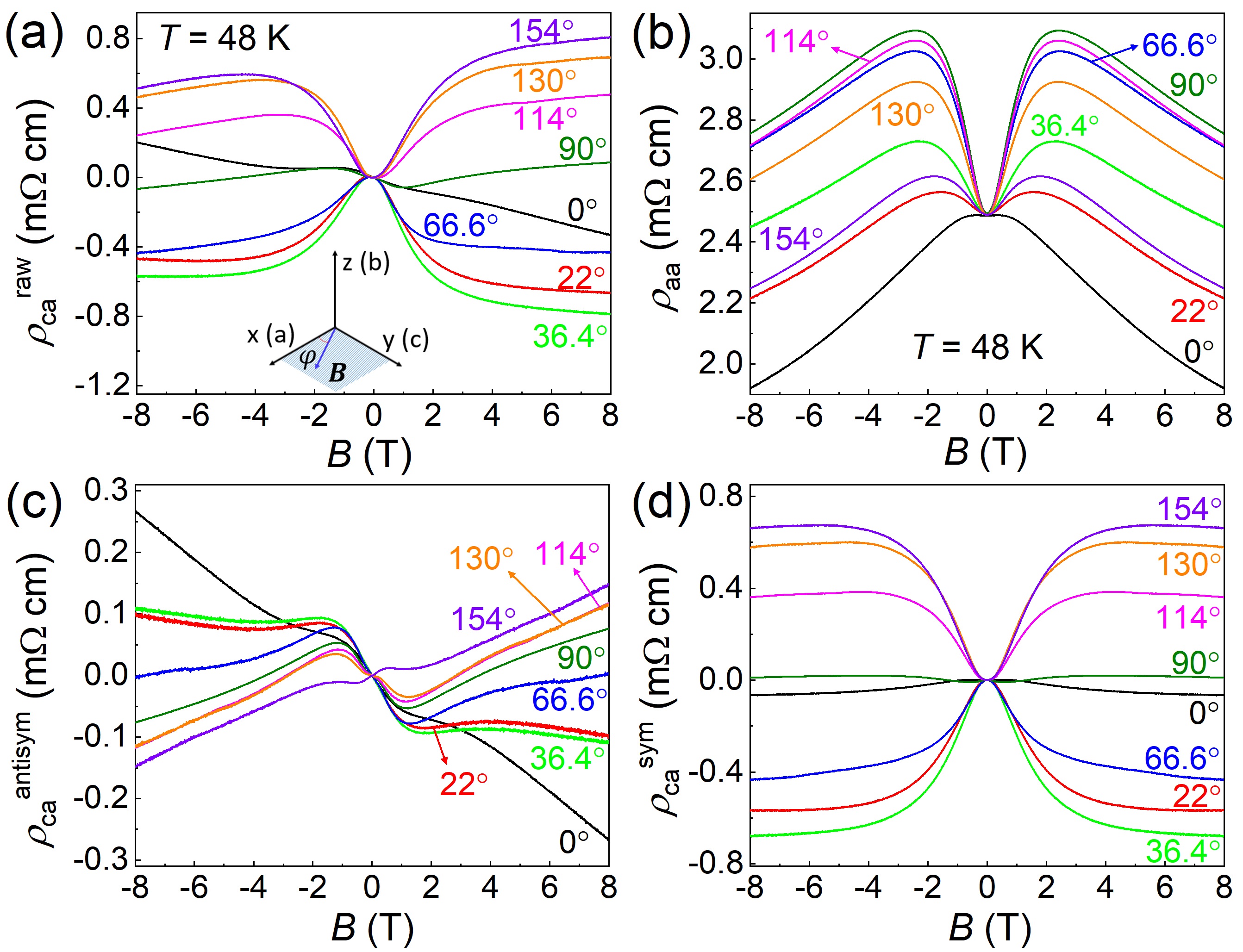}
	\caption{(a) $B$-dependences of the raw Hall resistivity $\rho_{ca}$ at various magnetic-field angles $\varphi$ in the $ac$ plane at 48 K. Inset shows the definition of $\varphi$. (b) $B$-dependences of the magnetoresistance at various $\varphi$. (c, d) $B$-dependences of the symmetric and antisymmetric components of $\rho_{ca}$ calculated from the data in (a) for the same set of $\varphi$ values.}
	\label{fig:2}
\end{figure}

Next, we studied the in-plane Hall effect, which may contain both PHE and  in-plane UHE, as a function of the magnetic-field angle $\varphi$ rotated in the $ac$ plane [see Fig.~2(a) inset]. For these measurements, similar to the one for $B \parallel a$-axis, for each fixed $\varphi$ we swept the magnetic field in a narrow range in the vertical plane across the $ac$ plane to find the alignment to guarantee $B_{\perp}$ = 0; namely, we measured both $\rho_{aa}$ and $\rho_{ca}$ during the sweep and chose the alignment that gives the minimum in $\rho_{aa}$ \cite{SM}.
The magnetic-field dependence of the raw $\rho_{ca}$ data at 48 K for various values of $\varphi$ in the $ac$ plane are shown in Fig. 2(a), where a small mixture of $\rho_{aa}$ due to the slight misalignment of the Hall-voltage contacts was removed. The magnetoresistance (MR), $\rho_{aa}(B)$, is shown for the same values of $\varphi$ in Fig. 2(b), where one can see that a negative MR is always present at high field above $\sim$3 T. The origin of this behavior, which cannot be simply due to chiral anomaly  \cite{Armitage2018}, is not understood yet. 

The antisymmetrized and symmetrized $B$-dependences of $\rho_{ca}$, which we call $\rho_{ca}^{\rm antisym}(B)$ and $\rho_{ca}^{\rm sym}(B)$, respectively, are shown in Figs. 2(c) and 2(d). Interestingly, $\rho_{ca}^{\rm antisym}(B)$ is the largest at 0$^{\circ}$ and shows a complicated sign-changing behavior at intermediate values of $\varphi$ [the plot of the $\varphi$-dependence of $\rho_{ca}^{\rm antisym}$ at $B$ = 8 T is shown in Fig. 3(a)]. 
On the other hand, $\rho_{ca}^{\rm sym}(B)$ is essentially zero at 0$^{\circ}$ and 90$^{\circ}$, and its magnitude at high-enough field obeys the $\sin (2\varphi)$ behavior expected for PHE [see Fig. 3(b)]. This result on $\rho_{ca}^{\rm sym}(B)$ is consistent with the PHE results reported for ZrTe$_5$ in Refs. \cite{Ge2020, Li2018}, where this was interpreted to be caused by the chiral anomaly \cite{Armitage2018}.  
The $\varphi$-dependence of the MR at 8 T is shown in Fig. 3(c), where one can see an essentially two-fold-symmetric behavior, which is similar to the behavior of PHE. The temperature dependence of the PFHE signal (i.e. $\rho_{ca}^{\rm antisym}$ for $B \parallel a$-axis) at 8 T shows a pronouced increase below $\sim 60$ K, which is probably related to the temperature dependence of the carrier density as we discuss later.

\begin{figure}[t]
	\centering
	\includegraphics[width=8.6cm]{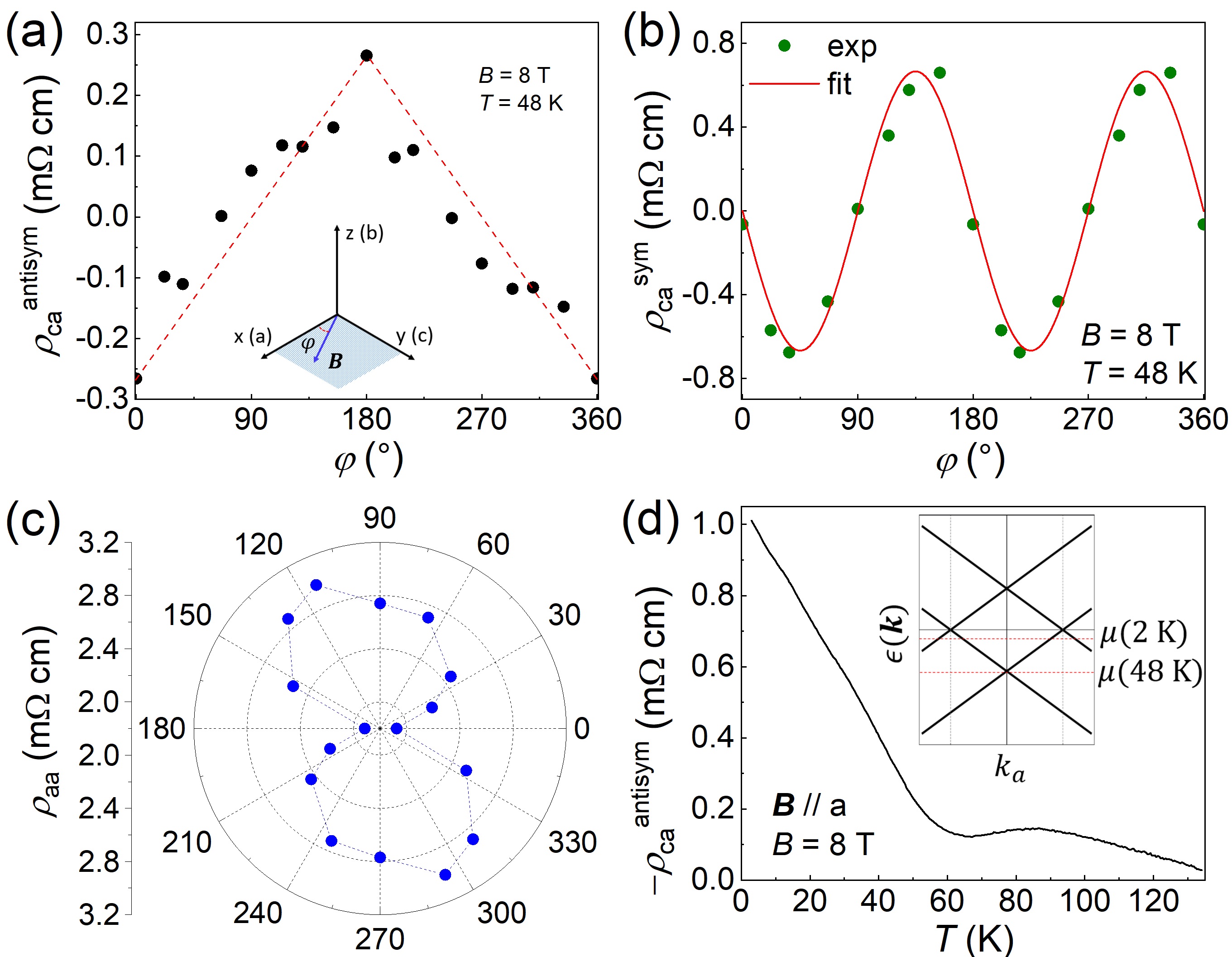}
	\caption{(a, b, c) $\varphi$-dependences of $\rho_{ca}^{\rm antisym}$, $\rho_{ca}^{\rm sym}$, and $\rho_{aa}$ at 8 T and 48 K, respectively. The data were measured from $0^{\circ}$ to $154^{\circ}$ (shown in Fig. 2) and reused for $180^{\circ}$ to $360^{\circ}$ because $B$ at $\varphi + 180^{\circ}$ is the same as $-B$ at $\varphi$. (d) Temperature dependence of $-\rho_{ca}^{\rm antisym}$ for $B \parallel a$-axis at 8 T. Inset shows the nodal-ring band structure of ZrTe$_5$ with $T_p$= 0 K, along with the location of the chemical potential at 2 K and 48 K.}
	\label{fig:3}
\end{figure}

As already mentioned, the existence of a finite PFHE indicates that the symmetry of ZrTe$_5$ must be lower than $Cm$, but the elucidation of the exact crystal structure would require a very high resolution analysis and difficult, given the easily deformable nature of the crystal. Nevertheless, we can quantitatively understand the observed PFHE by simply assuming the absence of any symmetry. If there is no symmetry constraint, the torus-shaped Fermi surface realized in ZrTe$_5$ under the $Cm$ symmetry \cite{Wang2022} can be tilted towards any direction, which makes the principal axes of the transport properties (along which the conductivity tensor is diagonalized) to be different from the crystallographic axes. 
This misalignment between the two coordinate systems can lead to finite off-diagonal components of the conductivity tensor in the crystallographic coordinate. In the following, we show that such an off-diagonal component can explain the observed PFHE in ZrTe$_5$.

Let us define two coordinate systems: the coordinate system $xyz$ defined by the eigenvectors of the conductivity tensor dictated by the Fermi-surface axes (eigen-coordinate) and the crystallographic coordinate system ($acb$). We make a simple assumption that the $acb$ and $xyz$ coodinates are related by a rotation matrix
\begin{equation}
\begin{split}
R(\alpha, \beta)=\left[\begin{array}{ccc}
\cos \beta & 0 & \sin \beta \\
\sin \alpha \sin \beta & \cos \alpha & -\sin \alpha \cos \beta \\
-\cos \alpha \sin \beta & \sin \alpha & \cos \alpha \cos \beta
\end{array}\right] \,\,,
\label{F1}
\end{split}
\end{equation}
where $\alpha$ and $\beta$ are the angles of clockwise rotation about $x$- and $y$-axes, respectively. This is motivated by the inference that a small tilting of the torus Fermi surface \cite{Wang2022} is the cause of the difference in the two coordinate systems.
We write the conductivity tensor in the eigen-coordinate system as 
\begin{equation}
\begin{split}
\sigma^{E}(B)=\left[\begin{array}{lll}
\sigma_{x x} & \sigma_{x y} & \sigma_{x z} \\
\sigma_{y x} & \sigma_{y y} & \sigma_{y z} \\
\sigma_{z x} & \sigma_{z y} & \sigma_{z z}
\end{array}\right] \,\,,
\label{F3}
\end{split}
\end{equation}
where all the off-diagonal components are zero in $B$ = 0.
The conductivity tensor in the crystallographic coordinate system
\begin{equation}
\begin{split}
\sigma^{C}(B)=\left[\begin{array}{lll}
\sigma_{a a} & \sigma_{a c} & \sigma_{a b} \\
\sigma_{c a} & \sigma_{c c} & \sigma_{c b} \\
\sigma_{b a} & \sigma_{b c} & \sigma_{b b}
\end{array}\right] \label{F5}
\end{split}
\end{equation}
is related to $\sigma^{E}(B)$ via
\begin{equation}
\begin{split}
\sigma^{C}(B)=R(\alpha, \beta) \sigma^{E}(B) R^{\dagger}(\alpha, \beta)  \,\,.
\label{F7}
\end{split}
\end{equation}

In ZrTe$_5$, the measurements of the Shubnikov-de Haas oscillations that elucidated the torus-shaped Fermi surface found that the possible misalignment of the Fermi-surface axes with respect to the crystallographic axes does not exceed 1$^{\circ}$ \cite{Wang2022}, which means $\alpha , \beta \leq$ 1$^{\circ}$. 
Hence, when a magnetic field is applied along the crystallographic $a$-axis, $\sigma_{xy}$, $\sigma_{yx}$, $\sigma_{xz}$, and $\sigma_{zx}$ in Eq.~\eqref{F3} would be very small and the approximation 
\begin{equation}
\begin{split}
\sigma^{E}(B) \approx \left[\begin{array}{lll}
\sigma_{x x} & 0 & 0 \\
0 & \sigma_{y y} & \sigma_{y z} \\
0 & -\sigma_{y z} & \sigma_{z z}
\end{array}\right] {\rm (when}\,\, B \parallel a{\rm -axis)}
\label{F8}
\end{split}
\end{equation}
 is justified.
From Eqs.~\eqref{F7} and \eqref{F8}, one obtains 
\footnotesize
\begin{eqnarray}
\sigma_{b a}  \approx&-\sigma_{x x} \cos \alpha \sin \beta \cos \beta+\sigma_{y z} \sin \alpha \sin \beta + \sigma_{z z} \cos \alpha \sin \beta \cos \beta \nonumber\\
\sigma_{c b}  \approx&-\frac{\sigma_{x x} \sin 2 \alpha \sin ^{2} \beta- \sigma_{y y} \sin 2 \alpha-2\sigma_{y z} \cos \beta+ \sigma_{z z} \sin 2 \alpha \cos ^{2} \beta }{2} \nonumber\\
\sigma_{b b}  \approx&\sigma_{x x} \cos ^{2} \alpha \sin ^{2} \beta+\sigma_{y y} \sin ^{2} \alpha+\sigma_{z z} \cos ^{2} \alpha \cos ^{2} \beta \nonumber\\
\sigma_{c a}  \approx&\sigma_{x x} \sin \alpha \sin \beta \cos \beta+\sigma_{y z} \cos \alpha \sin \beta- \sigma_{z z} \sin \alpha \sin \beta \cos \beta \nonumber\\
\label{F9}
\end{eqnarray}
\normalsize
for $B \parallel a$-axis.
Since our purpose is to explain the finite PFHE for $B \parallel a$-axis, let us focus on $\rho_{ca}$. Since the resistivity tensor is the inverse of the conductivity tensor, i.e. $\rho^{C}(B)=\left[\sigma^{C}(B)\right]^{-1}$, we have
\begin{equation}
\begin{split}
\rho_{c a} = \left(\sigma_{b a} \sigma_{c b}-\sigma_{b b} \sigma_{c a}\right) / D \,\,,
\end{split}  \label{F10}
\end{equation}
with the determinant $D$ given by
\footnotesize
\begin{eqnarray}
 D=\left|\begin{array}{lll}
\sigma_{a a} & \sigma_{a c} & \sigma_{a b} \\
\sigma_{c a} & \sigma_{c c} & \sigma_{c b} \\
\sigma_{b a} & \sigma_{b c} & \sigma_{b b}
\end{array}\right| \approx& \left|\begin{array}{ccc}
\sigma_{x x} & 0 & 0 \\
0 & \sigma_{y y} & \sigma_{y z} \\
0 & -\sigma_{y z} & \sigma_{z z} 
\end{array}\right| 
= \sigma_{x x}\left(\sigma_{y y} \sigma_{z z}+\sigma_{y z}^{2}\right) \,. \nonumber\\
\label{F11}
\end{eqnarray}
\normalsize
 
We can thus calculate $\rho_{ca}$ based on the inputs of $\alpha$, $\beta$, $\sigma_{xx}$, $\sigma_{yy}$, $\sigma_{zz}$, and $\sigma_{yz}$, which leads to
\begin{equation}
\rho_{c a} (B)\approx \frac{- \sigma_{y z} \sin \beta - \sigma_{yy} \sin \alpha \sin \beta}{\left(\sigma_{yy} \sigma_{z z}+\sigma_{y z}^{2}\right)} + \frac{\sin \alpha \sin \beta}{\sigma_{xx}} \,,
\label{F12}
\end{equation}
where we assumed $\sin^2 \alpha \approx \sin^2 \beta \approx 0$ and $\cos \alpha \approx \cos \beta \approx 1$.

Since $\sigma_{yz}$ and $\sigma_{ii}$ are antisymmetric and symmetric with $B$, respectively, $\rho_{ca}(B)$ has both antisymmetric and symmetric components. They can be separately written as
\begin{equation}
\rho_{c a}^{\rm antisym} (B)\approx \frac{- \sigma_{y z} \sin \beta }{\left(\sigma_{yy} \sigma_{z z}+\sigma_{y z}^{2}\right)} \\
\label{F13}
\end{equation}
and
\begin{equation}
\rho_{c a}^{\rm sym} (B)\approx \left[\frac{- \sigma_{yy} }{\left(\sigma_{yy} \sigma_{z z}+\sigma_{y z}^{2}\right)} + \frac{1}{\sigma_{xx}}\right]\sin \alpha \sin \beta \,. \\
\label{F14}
\end{equation}

One can see from Eq.~\eqref{F13} that the antisymmetric PFHE for $B \parallel a$-axis comes from the conventional Hall conductivity $\sigma_{yz}$ due to the coordinate rotation $\beta$. One can also see from Eq.~\eqref{F14} that $\rho_{c a}^{\rm sym} (B)\approx$ 0, because $\alpha , \beta \leq$ 1$^{\circ}$; this is in agreement with the experimental observation. Using the general relation $\rho_{y z}= -\sigma_{y z}/(\sigma_{y y} \sigma_{z z}+\sigma_{y z}^{2})$, Eq.~\eqref{F13} reduces to
\begin{equation}
\rho_{c a}^{\rm antisym} (B)\approx \rho_{yz}(B) \sin \beta \,.
\label{F15}
\end{equation}
Therefore, to estimate the actual size of $\rho_{c a}^{\rm antisym}$, we just need the values of $\rho_{yz}$ and $\beta$.
Since it is very difficult to reliably measure $\rho_{yz}$ on a soft ZrTe$_5$ flakes and there are actually no $\rho_{yz}$ data available in the literature,
we make a crude assumption that $\rho_{yz}$ is simply given by the carrier density $n$, which we calculate from a single-band model. Note that we do not use the literature data on $\rho_{yx}$ to calculate $n$, because the small quantum-limiting field and the large anomalous Hall effect both make it difficult to extract $n$ from $\rho_{yx}(B)$ \cite{Liang2018, Wang2022}. 
The angle-resolved photoemission spectroscopy (ARPES) results show a temperature-dependent chemical potential $\mu$ in ZrTe$_5$ \cite{YZhang2017, Fu2020}, and $\mu$ shifts by $\sim$15 meV when the temperature is 46 K away from the $T_p$ temperature. For the samples with $T_p$ = 0 K, $\mu$ is estimated to be $\sim$5 meV at 2 K, if we use $n = 2.3 \times 10^{16}$ cm$^{-3}$ obtained from quantum oscillations \cite{Wang2022}. Thus, one can estimate $\mu$ to be $\sim$20 meV at 48 K; this is close to the inversion-symmetry breaking energy (19.1 meV), but one can still crudely treat the Fermi surface as a torus, for which $n$ is proportional to $\mu^2$ \cite{Wang2022}. This leads to the estimation $n \approx 3.7 \times 10^{17}$ cm$^{-3}$ at 48 K, which would correspond to $\rho_{yz} \approx$ 13.5 m$\Omega$cm at 8 T. 
Assuming $\beta = -1^{\circ}$, one arrives at the estimate $\rho_{ca} \approx -0.236$ m$\Omega$cm for $B \parallel a$-axis of 8 T, which is close to our experimental result, $-0.266$ m$\Omega$cm. Therefore, the finite antisymmetric PFHE observed in ZrTe$_5$ can be quantitatively understood by assuming a tilting of the long axis of the torus Fermi surface (which is approximately along the $b$-axis) towards the $a$-axis by only $-1^{\circ}$. Since $\mu$ moves closer to the Dirac point at lower temperature to make $n$ smaller, one would expect $\rho_{yz}$, and hence $\rho_{ca}$, to increase with lowering temperature, which is consistent with the result in Fig. 3(d). 

In conclusion, by carefully eliminating the effect of out-of-plane magnetic field, we have elucidated the existence of a finite parallel-field Hall effect in ZrTe$_5$. To the best of our knowledge, this is the first case that the existence of the parallel-field Hall effect is unambiguously established in any material. Its origin is however not related to topology, but it can be understood as a result of very low symmetry of the system. Therefore, our result demonstrates that the parallel-field Hall effect is a powerful indicator of symmetry breaking in a material, which is useful because subtle symmetry breaking is often difficult to be nailed down by structural analyses \cite{Wang2022}. Besides, this study provides a useful example that special care must be taken upon analyzing the unconventional Hall effect in a topological material when the symmetry of the system is low. This is particularly important because a finite integrated Berry curvature always requires some symmetry breaking \cite{Nagaosa2010}. Careful symmetry analysis as performed in Ref. \cite{Wang2024} should be the starting point of understanding the unconventional Hall effect.

The raw data used in the generation of figures are available at the online repository Zenodo \cite{Zenodo}.

\acknowledgements{This work has received funding from the Deutsche Forschungsgemeinschaft (DFG, German Research Foundation) under CRC 1238-277146847 (subprojects A04, B01, and C02) and also from the DFG under Germany’s Excellence Strategy -- Cluster of Excellence Matter and Light for Quantum Computing (ML4Q) EXC 2004/1-390534769. }

\end{document}


{\bf{\large Supplemental Material for ``Parallel-Field Hall effect in ZrTe$_5$"}
}

\vspace{5mm}

\subsection{Method}

Single crystals of ZrTe$_5$ with $T_{\rm p}$ = 0 K were grown by Te-flux method. The details were reported in Ref. \cite{Wang2022}. ZrTe$_5$ flakes were mechanically exfoliated from high quality single crystals and tranferred to a wafer (SiO$_2$ 290 nm on doped Si). The brief fabrication process of the electrodes is shown below:\\
(1) Optical lithography by Heidelberg µPG 101 Laser Writer was performed after the spin-coating of AZ1505 photoresist \\
(2) Development of the pattern in AZ326 developer\\
(3) In-situ cleaning by Ar plasma, followed by sputter-deposition of 10 nm Au / 450 nm Nb film\\
(4) Lift-off in acetone \\
Transport measurements were performed in a Quantum Design Physical Properties Measurement System (PPMS) with a rotating sample holder. The resistance and the Hall resistance were measured simultaneously by four-terminal method using a low-frequency (17.33 Hz) AC lock-in technique.

\subsection{$\theta$ dependences of the magentoresistance and $\rho_{ca}$ for various $\varphi$}

\begin{figure}[h]
	\centering
	\includegraphics[width=13cm]{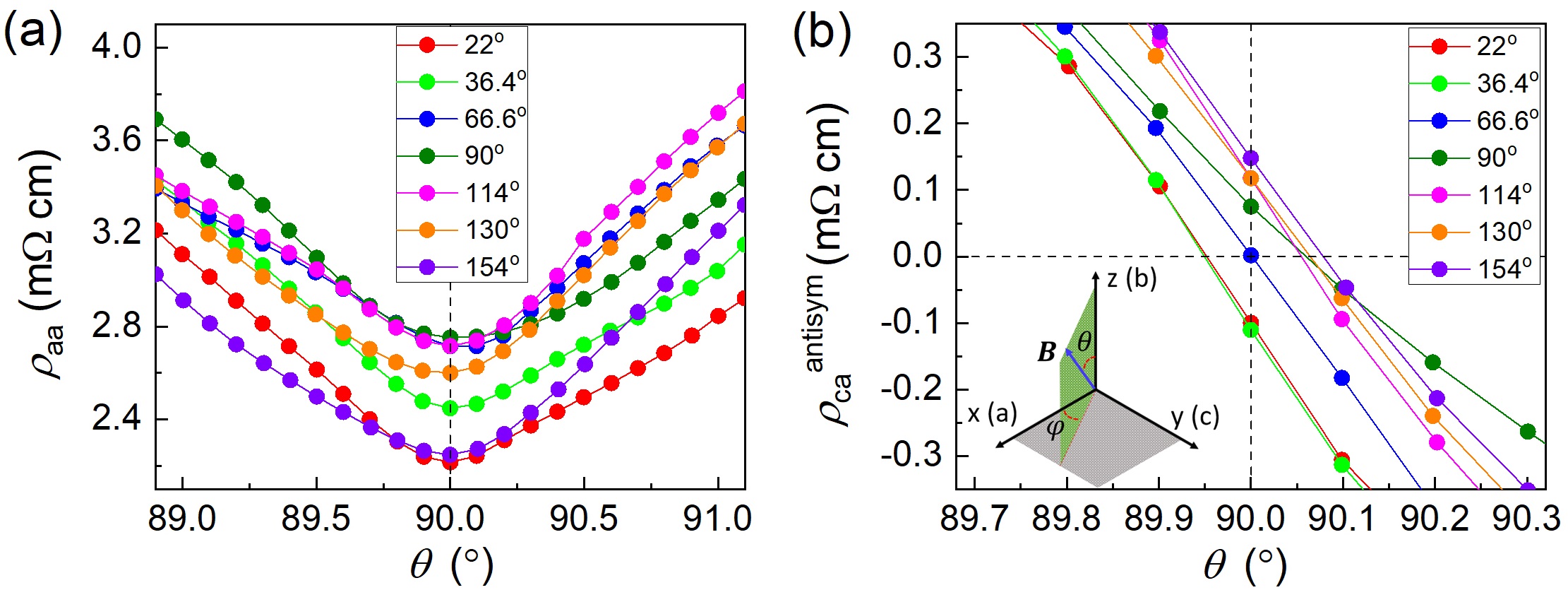}
	\caption{(a-b) Dependence of magnetoresistance and antisymmtrized Hall effect on magnetic-field angle $\theta$ in the vertical plane at 8 T and 48 K for various angles $\varphi$ in $ac$ plane. Inset in (b) shows the definition of angle $\theta$ and $\varphi$. }
	\label{fig:S1}
\end{figure}

Fig. S1 shows the dependences of symmetrized magnetoresistance and antisymmtrized in-plane Hall resistance on the magnetic-field angle $\theta$ swept in the vertical plane at 8 T and 48 K, measured for various angle $\varphi$ changed in the horizontal $ac$ plane. The angle $\theta$, at which the magnetoresistance shows the minimum [see Fig. S1(a)], was defined to be 90$^{\circ}$ for each $\varphi$. The inset in Fig. S1(b) shows the definition of the angles $\theta$ and $\varphi$. The $\theta$-dependence of the in-plane Hall resistivity shows a nonzero value at 90$^{\circ}$ at all values of $\varphi$ except for 66.6$^{\circ}$, as shown in Fig. S1(b).

\subsection{Comparison between the present result and Ref. \cite{Ge2020} by Ge {\it et al.}}

\begin{figure}[h]
	\centering
	\includegraphics[width=8cm]{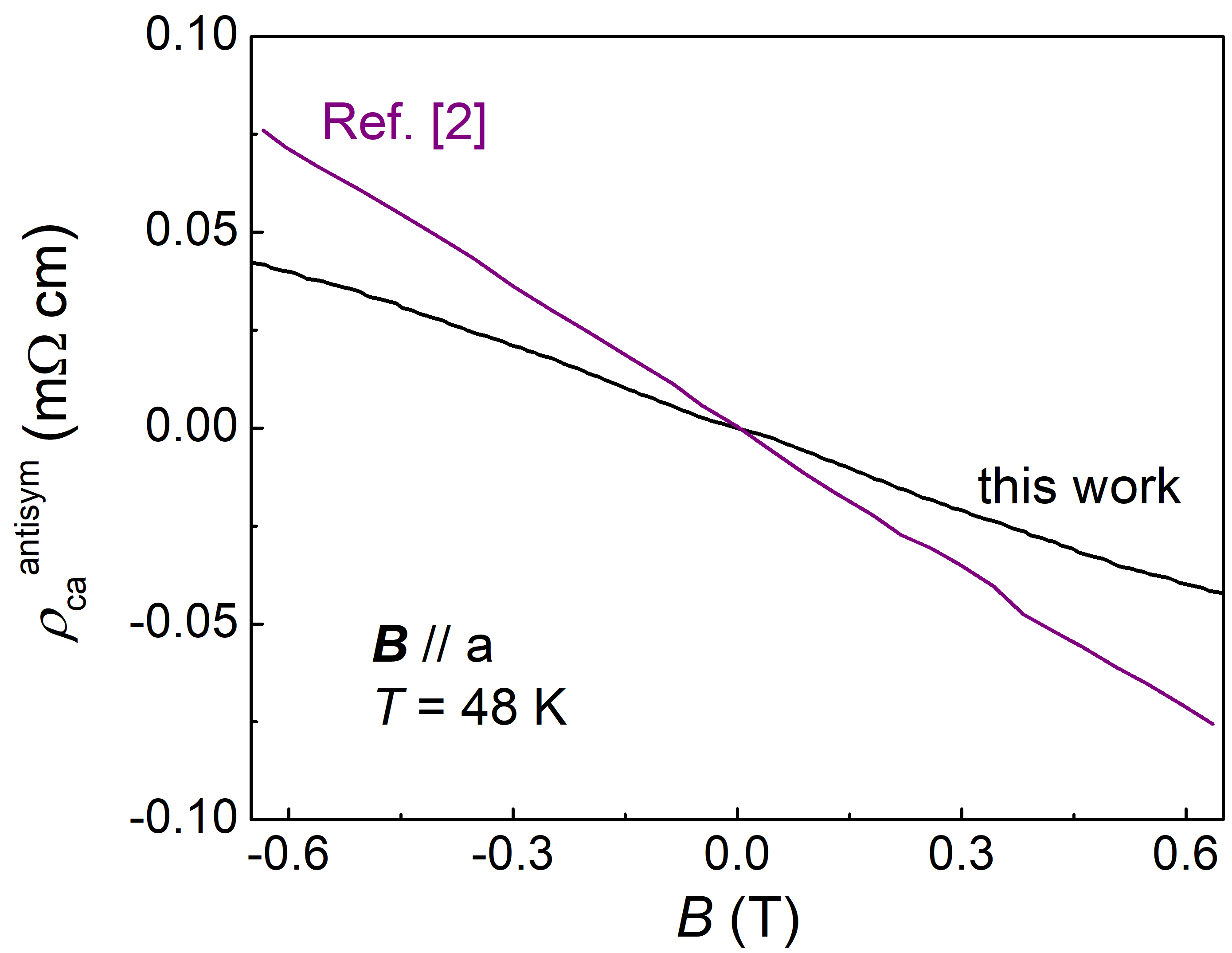}
	\caption{Magnetic-field dependence of the antisymmetric component of $\rho_{ca}$ for $B \parallel a$-axis in the low-field range at 48 K. Black and purple curves are from this work and Ref. \cite{Ge2020}, respectively. } \label{fig:S2}
\end{figure}

Different from the ZrTe$_5$ samples with $T_p \simeq$ 60 K used in Ref. \cite{Ge2020}, the $T_p$ = 0 K samples used in the present work has a very low carrier density at low temperature, and it is not reasonable to compare the low-temperature results between samples with different $T_p$ values. If we check the temperature dependence of the resistivity of our sample and the main sample (S2) in Ref. \cite{Ge2020}, we can see that the resistivity values at $\sim$50 K ($\sim$2.4 m$\Omega$cm) are roughly the same for both samples.  Hence, we used the results at $\sim$50 K for the comparison between the samples used in this work and in Ref. \cite{Ge2020}. As shown in Fig. S2, the value of the antisymmetric Hall signal from Ref. \cite{Ge2020} for $B \parallel a$-axis at 50 K and $-0.6$ T is around 0.075 m$\Omega$cm, which is larger than our results of 0.04 m$\Omega$cm. For higher magnetic field at $-8$ T, our results for $B \parallel a$-axis at 48 K is around 0.26 m$\Omega$cm as shown in the main text, which is still smaller than the value of around 0.5 m$\Omega$cm at 50 K in Ref. \cite{Ge2020}. Our results show smaller values of the antisymmetric in-plane Hall effect for $B \parallel a$-axis compared to the results in Ref. \cite{Ge2020}, which suggests that our result is more free from the possible contamination due to $B_{\perp}$.

\subsection{Estimation of $\rho_{ca}$ in zero magnetic field}

In zero magnetic field, Eq. (9) in the main text reduces to
\begin{equation}
\rho_{c a} \approx (\rho_{x x} - \rho_{z z}) \sin \alpha \sin \beta
\label{F12}
\end{equation}
because $\sigma_{y z} = 0$.
From the literature, the resistivity anisotropy in ZrTe$_5$ can be estimated to be $\rho_{z z} \approx 50\rho_{x x}$ at the $T_{\rm p}$ temperature \cite{Lv2017}, which leads to $\rho_{c a} \approx$ 0.036 m$\Omega$cm with the assumption of $\alpha = -\beta = 1^{\circ}$ and the input of $\rho_{x x} \approx $ 2.4 m$\Omega$cm. This $\rho_{c a}$ is one order of magnitude smaller than the parallel-field Hall effect at 8 T ($-0.266$ m$\Omega$m).